\documentclass{ifacconf}

\usepackage{graphicx}      
\usepackage{natbib}        
\usepackage{pdfsync}
\usepackage{epsfig} 
\usepackage{cuted}
\usepackage{lipsum}
\usepackage{mathtools}
\usepackage{amsmath,amssymb,amsfonts}
\usepackage{mathrsfs}
\usepackage{enumerate}
\usepackage{color}
\usepackage{subfigure}
\usepackage{comment}
\usepackage{algorithm}
\usepackage{flushend}
\usepackage{algpseudocode}

\newtheorem{mydef}{Definition}
\newtheorem{mythm}{Theorem}

\newtheorem{remark}{Remark}
\begin{document}
\begin{frontmatter}

\title{Linear Data-Driven Economic MPC with Generalized Terminal Constraint\thanksref{footnoteinfo}}

\thanks[footnoteinfo]{F. Allg\"{o}wer is thankful that his work was funded by Deutsche Forschungsgemeinschaft (DFG, German Research
Foundation) under Germany’s Excellence Strategy - EXC 2075 - 390740016 and under grant 468094890.
F. Allg\"{o}wer acknowledge the support by the Stuttgart
Center for Simulation Science (SimTech).
The authors thank the International Max Planck Research School
for Intelligent Systems (IMPRS-IS) for supporting Yifan Xie.}

\author[First]{Yifan Xie}
\author[First]{Julian Berberich}
\author[First]{Frank Allg\"{o}wer}

\address[First]{University of Stuttgart, Institute for Systems Theory and Automatic Control, 70550 Stuttgart, Germany (e-mail: \{yifan.xie, julian.berberich, frank.allgower\}@ist.uni-stuttgart.de).}

\begin{abstract}                
In this paper, we propose a data-driven economic model predictive control (EMPC) scheme with generalized terminal constraint to control an unknown linear time-invariant system.
Our scheme is based on the Fundamental Lemma to predict future system trajectories using a persistently exciting input-output trajectory.
The control objective is to minimize an economic cost objective.
By employing a generalized terminal constraint with artificial equilibrium, the scheme does not require prior knowledge of the optimal equilibrium.
We prove that the asymptotic average performance of the closed-loop system can be made arbitrarily close to that of the optimal equilibrium.
Moreover, we extend our results to the case of an unknown linear stage cost function, where the Fundamental lemma is used to predict the stage cost directly.
The effectiveness of the proposed scheme is shown by a numerical example.
\end{abstract}

\end{frontmatter}

\section{Introduction}\label{intro}

Data-driven system analysis and control have recently received increasing interest in the control community.
Data-driven methods are beneficial when it is hard to determine the system model and design a model-based controller.
For a linear time-invariant (LTI) system,
all trajectories can be constructed based on a persistently exciting data trajectory by employing the Fundamental Lemma proposed by \cite{willems2005note} in the behavioral approach.

Recently, several works have developed data-driven model predictive control (MPC) approaches based on \cite{willems2005note}, e.g., \cite{yang2015reduced} and \cite{coulson2019data}.
\cite{berberich2021guarantees} provide a theoretical analysis of closed-loop properties of the data-driven MPC scheme and also in case of noisy data.
Further approaches in this direction are summarized in the recent survey by \cite{markovsky2021reviews}.
Existing data-driven MPC schemes typically consider a standard tracking MPC that drives the system's input and output to a given setpoint.
In various practical fields such as process industries and transportation, the objective is to operate the system to achieve the best economic performance, which depends on the production goals, prices in the market, etc.
The system may not attain the best economic performance at any (known) steady setpoint.
In this case, rather than tracking the system trajectory to a prescribed setpoint, one wants to minimize an economic objective that has more than one minimum point.

Economic model predictive control (EMPC) addresses precisely this control problem in the context of model-based MPC \citep{faulwasser2018enmpc}.
\cite{diehl2011empc} develop an EMPC scheme to stabilize the optimal steady state for a given economic cost function based on a terminal equality constraint.
\cite{Amrit2011economicOU} propose a terminal region constraint instead of a point constraint in the EMPC scheme.
In \cite{fagiano2013generalized}, the authors consider a generalized terminal state constraint to provide a larger region of attraction by only requiring the terminal state to be \emph{some} steady state rather than the optimal one, see also \cite{ferramosca2014changing} and \cite{muller2013tuning}.
The above-mentioned works all design control strategies with the knowledge of the system model.
In practice, one often only has access to input-output data samples and it may be difficult to determine an accurate model of the system.

In this paper, we propose a data-driven EMPC scheme for LTI systems using a generalized terminal constraint.
The scheme only relies on a persistently exciting trajectory of an unknown system, using the Fundamental Lemma to predict further trajectories.
The generalized terminal constraint requires that the terminal setpoint is any equilibrium rather than the optimal one.
This idea is similar to the terminal constraints in \cite{fagiano2013generalized} and \cite{berberich2020changing}, which consider model-based EMPC and data-driven tracking MPC, respectively.
In contrast to existing data-driven MPC approaches such as \cite{berberich2020changing}, the implementation does not require a priori knowledge of the optimal equilibrium, which can be particularly hard to obtain in a data-driven control setup.
We prove recursive feasibility, constraint satisfaction and asymptotic average performance guarantees for the proposed data-driven EMPC.
Finally, we propose a novel data-driven EMPC scheme that can also cope with unknown (linear) cost functions, which is a relevant problem in the recent learning-based control literature, e.g., \cite{nonhoff2022online, manzano2021oracle, gros2020data}.
In conclusion, the proposed data-driven EMPC scheme has additional advantage to deal with unknown (linear) cost functions compared with the model-based EMPC scheme.

The remainder of this paper is organized as follows. In Section~\ref{pre}, necessary preliminaries are introduced.
In Section~\ref{scheme}, we first propose the data-driven EMPC scheme with generalized terminal constraints.
Then, we show that the scheme ensures that the asymptotic average performance of the closed-loop system converges close to the optimal performance.
In Section~\ref{unknown}, we present the data-driven EMPC scheme with unknown and linear cost function.
We apply the developed scheme to a numerical example in Section~\ref{simulation}.
Finally, we conclude the paper in Section~\ref{conclusion}.

\section{Preliminaries}\label{pre}
Let $\mathbb{I}_{[a, b]}$ denote the set of integers in the interval $[a, b]$ and $\mathbb{I}_{\geq 0}$ denote the set of nonnegative integers.
Given a sequence $\{x_k\}_{k=0}^{N-1}$ with $x_k\in\mathbb{R}^n$, we define $x_{[a, b]}:=[x_a^\top \ldots x_b^\top]^\top\in\mathbb{R}^{(b-a+1)n}$ and $x:=x_{[0, N-1]}\in\mathbb{R}^{Nn}$.
A Hankel matrix is defined as
\begin{equation}\label{hankel}
H_L(x):=
\left[
  \begin{array}{cccc}
    x_0 & x_1 &\ldots & x_{N-L}\\
    x_1 & x_2 &\ldots & x_{N-L+1}\\
    \vdots &\vdots &\ddots &\vdots\\
    x_{L-1} &x_L &\ldots &x_{N-1}
  \end{array}\nonumber
\right].
\end{equation}

\subsection{Data-driven system representation}
In this paper, we consider an unknown LTI system in the form of
\begin{equation}\label{system}
\begin{aligned}
x_{k+1}&=Ax_k+Bu_k,\\
y_{k}&=Cx_k+Du_k,
\end{aligned}
\end{equation}
where $x_k\in \mathbb{R}^n$ denotes the state, $u_k\in \mathbb{R}^m$ denotes the input, and $y_k\in \mathbb{R}^{p}$ denotes the output.
We assume that the matrices $(A, B, C, D)$ are a minimal realization of the system \eqref{system}, which means that $(A, B)$ is controllable and $(A, C)$ is observable.
Moreover, the matrices $(A, B, C, D)$ are unknown, but one input-output trajectory $\{u_k^d, y_k^d\}_{k=0}^{N-1}$ of system \eqref{system} is available.
We consider the standard definition of persistence of excitation, compare \cite{willems2005note}.

\begin{mydef}\upshape
A sequence $\{u_k\}_{k=0}^{N-1}$ with $u_k\in \mathbb{R}^m$ is persistently exciting of order $L$ if $\text{rank}(H_L(u))=mL$.
\end{mydef}
This definition means that a input sequence is sufficiently rich.
We now introduce the Fundamental Lemma which was first proposed in the context of behavioral systems theory by \cite{willems2005note}.

\begin{mythm}\label{FL}\upshape
Suppose that $\{u_k^d, y_k^d\}_{k=0}^{N-1}$ is an input-output trajectory of the LTI system \eqref{system} and $u^d$ is persistently exciting of order $L+n$. Then, a sequence $\{u_k, y_k\}_{k=0}^{L-1}$ is an input-output trajectory of system \eqref{system} if and only if there exists a vector $\alpha\in \mathbb{R}^{N-L+1}$ such that
\begin{equation}
\begin{bmatrix}
    H_L(u^d)\\
    H_L(y^d)
\end{bmatrix}\alpha
=
\begin{bmatrix}
u\\
y
\end{bmatrix}.\nonumber
\end{equation}
\end{mythm}
By Theorem \ref{FL}, a finite-length input-output trajectory generated by a persistently exciting input sequence can represent the behavior of the system.
This allows us to reconstruct further system trajectories without access to a state-space model.
We will use Theorem \ref{FL} to set up a data-driven EMPC scheme in the remainder of the paper.

\subsection{Problem setup}
In this paper, we use data-driven EMPC to minimize a general, economic goal, while the inputs and outputs satisfy the constraints.
We consider pointwise-in-time constraints on the input and output signals in the form $u_t\in \mathbb{U}$ and $y_t\in \mathbb{Y}$ for a compact set $\mathbb{U}\subseteq \mathbb{R}^m$  and a closed set $\mathbb{Y}\subseteq \mathbb{R}^p$.
The economic stage cost function is defined as
\[l:\mathbb{U}\times\mathbb{Y}\rightarrow\mathbb{R},\]
which is a scalar-valued continuous function that is minimized in the considered EMPC scheme.
As is common in EMPC, we do not impose further assumptions on convexity or positive (semi-)definiteness of $l$.

Since we only have access to input-output measurements, we define the equilibrium of the system via input-output pairs rather than states.
\begin{mydef}\upshape
An input-output pair $(u^e, y^e)\in\mathbb{U}\times\mathbb{Y}$ is an equilibrium of the system \eqref{system}, if the sequence $\{u_k, y_k\}_{k=0}^n$ with $(u_k, y_k)=(u^e, y^e)$ for all $k\in\mathbb{I}_{[0, n]}$ is a trajectory of system \eqref{system}.
The set of all equilibria of the system is denoted by $\mathbb{U}^e\times\mathbb{Y}^e\subseteq \mathbb{U}\times\mathbb{Y}$.
\end{mydef}

Next, we define the optimal equilibrium regarding the stage cost function $l$.
\begin{mydef}\upshape
An input-output pair $(u^{oe}, y^{oe})\in\mathbb{U}\times\mathbb{Y}$ is an optimal equilibrium of the system \eqref{system}, if
\[l(u^{oe}, y^{oe})=\min_{(u, y)\in \mathbb{U}^e\times\mathbb{Y}^e}l(u, y).\]
The set of all optimal equilibria is denoted by $\mathbb{U}^{oe}\times\mathbb{Y}^{oe}\subseteq \mathbb{U}^e\times\mathbb{Y}^e$.
\end{mydef}
Given an equilibrium $(u^{e}, y^{e})$, we define $u_n^{e}$ and $y_n^{e}$ as the column vectors containing $n$ times $u^{e}$ and $y^{e}$, respectively.

\section{Data-driven EMPC}\label{scheme}
We first present the data-driven EMPC optimization problem with generalized terminal constraint in Section~\ref{3.1}.
In Section~\ref{3.2}, we propose an algorithm to solve the optimization problem in a receding horizon manner.
We prove recursive feasibility, constraint satisfaction and economic performance guarantees for the proposed scheme in Section~\ref{clp}.

\subsection{EMPC optimization problem}\label{3.1}
At time $t$, we have an input-output trajectory  $\{u_k^d,
y_k^d\}_{k=0}^{N-1}$, initial conditions $\{u_k, y_k\}_{k=t-n}^{t-1}$ and a bound $\bar{l}(t)\geq l(u^{oe}, y^{oe})$.
Based on these ingredients, the data-driven EMPC scheme with generalized terminal constraint is formulated as follows:
\begin{subequations}\label{optprob1}
\begin{align}
&J_L^*(u_{[t-n, t-1]}, y_{[t-n, t-1]})=\nonumber\\
\min_{\alpha(t), u(t), y(t)\atop u^e(t), y^e(t)}&\sum_{k=0}^{L-1}l(u_k(t), y_k(t))+
\beta l(u^e(t), y^e(t))\label{ob1}\\
\text{s.t. }\quad &
\begin{bmatrix}
    u_{[-n, L]}(t)\\
    y_{[-n, L]}(t)
\end{bmatrix}=
\begin{bmatrix}
    H_{L+n+1}(u^d)\\
    H_{L+n+1}(y^d)
\end{bmatrix}\alpha(t),\label{cs1}
\\
&\label{cs2}
\begin{bmatrix}
u_{[-n, -1]}(t)\\
y_{[-n, -1]}(t)
\end{bmatrix}=
\begin{bmatrix}
u_{[t-n, t-1]}\\
y_{[t-n, t-1]}
\end{bmatrix},
\\
&\begin{bmatrix}\label{cs3}
    u_{[L-n, L]}(t)\\
    y_{[L-n, L]}(t)
\end{bmatrix}=
\begin{bmatrix}
    u_{n+1}^e(t)\\
    y_{n+1}^e(t)
\end{bmatrix},
\\
&(u_k(t), y_k(t))\in \mathbb{U}\times\mathbb{Y}, \forall k\in\mathbb{I}_{[0, L]},\label{cs4}\\
&l(u^e(t), y^e(t))\leq \bar{l}(t),\label{cs5}
\end{align}
\end{subequations}
where $L\in\mathbb{N}$ is a finite prediction horizon and $\beta\in \mathbb{R}^+$ is a scalar.
In constraint \eqref{cs1}, the input-output trajectory is parameterized by Theorem \ref{FL} based on data $\{u_k^d,
y_k^d\}_{k=0}^{N-1}$, where $u^d$ is persistently exciting of order $L+2n+1$.
In constraint \eqref{cs2}, we initialize the trajectory at time $t$ using past $n$ steps of the input-output trajectory $\{u_k, y_k\}_{k=t-n}^{t-1}$.
The terminal equality constraint \eqref{cs3} ensures that the internal state of the trajectory after $L$ steps is equal to the artificial steady-state $x^e(t)$ corresponding to the artificial equilibrium $(u^e(t), y^e(t))$.
The input and output constraints are included via \eqref{cs4}.
Similar to \cite{fagiano2013generalized}, the stage cost of the artificial equilibrium is required to be less than or equal to a given bound $\bar{l}(t)$ as will be explained later.
An optimal solution of problem \eqref{optprob1} is denoted by $\alpha^*(t), u^*(t)$, $y^*(t)$, $u^{e*}(t)$ and $y^{e*}(t)$.

Problem \eqref{optprob1} uses an artificial equilibrium to ensure a larger feasible region of the problem if compared to EMPC with a fixed equilibrium $(u^e(t),y^e(t))=(u^{oe},y^{oe})$.
The cost function contains the sum of the economic cost over $L$ steps in the future as well as the penalty $\beta l(u^e(t), y^e(t))$.
By adding this penalty, the performance of the closed-loop system  obtained in the receding-horizon fashion can be made arbitrarily close to that of the optimal equilibrium, which will be shown in Section~\ref{clp}.
Constraint \eqref{cs5} and a suitable update of $\bar{l}(t)$ (Algorithm 1, line 4) force the terminal stage cost to be not increasing with time $t$.

 Note that the proposed data-driven EMPC scheme does not require prior knowledge of the optimal equilibrium for its implementation, in contrast to the existing data-driven tracking MPC scheme by \cite{berberich2020changing},

In constraint \eqref{cs2}, the initial conditions can be written as an extended state
\begin{equation}\label{system2}
\begin{aligned}
\xi_t:=
\begin{bmatrix}
u_{[t-n, t-1]}\\
y_{[t-n, t-1]}
\end{bmatrix},
\end{aligned}\nonumber
\end{equation}
compare \cite[Lemma 2]{koch2022dddissipativity}.
The feasibility set of problem \eqref{optprob1} is defined by
\begin{equation}
\mathcal{F}:=
\left\{\xi_t:
\begin{gathered}
\text{there exists }\bar{l}
\text{ such that \eqref{optprob1} is feasible}
\end{gathered}
\right\},\nonumber
\end{equation}
which includes all initial conditions such that problem \eqref{optprob1} admits a solution.

Problem \eqref{optprob1} is similar to the MPC approach with generalized terminal constraint in \cite{fagiano2013generalized}.
The difference is that an input-output cost function is considered in our problem and we predict the input-output trajectories based on Hankel matrices.
When the stage cost $l$ is convex, the proposed scheme in \eqref{optprob1} is analogous to the data-driven MPC scheme with changing setpoints in \cite{berberich2020changing}.
Problem \eqref{optprob1} reduces to a data-driven EMPC problem with terminal equality constraint by adding a constraint
$(u^e(t), y^e(t))=(u^{oe}, y^{oe})$, leading to a data-driven version of the MPC scheme in \cite{diehl2011empc}.
The feasibility set of this data-driven EMPC problem with terminal equality constraint is defined by
\begin{equation}
\mathcal{F}^s:=
\left\{\xi_t:\!
\begin{gathered}
\text{problem \eqref{optprob1} with the additional constraint } \\
(u^e(t), y^e(t))=(u^{oe}, y^{oe}) \text{ admits a solution}
\end{gathered}
\right\},\nonumber
\end{equation}
which is used for closed-loop analysis in Section~\ref{clp}.

\subsection{Receding-horizon algorithm}\label{3.2}
Before we propose the data-driven EMPC scheme based on \eqref{optprob1}, two scalars $\beta$ and $\bar{l}(t)$ need to be chosen.
The scalar $\beta$ can be an arbitrary value that is larger than $0$ although its precise value influences the guaranteed closed-loop properties, see Theorem 3.
To choose the terminal cost bound $\bar{l}(t)$ at the initial time $t=0$, we define the set of reachable equilibria and the optimal achievable stage cost.
\begin{mydef}\upshape
Given an input-output trajectory  $\{u_k^d, y_k^d\}_{k=0}^{N-1}$ that is generated by a persistently exciting input sequence of order $L+2n+1$, initial conditions $\xi_t$, and a finite prediction horizon $L\in\mathbb{N}$, the set of reachable equilibria in $L$ steps is defined as
\begin{equation}
\mathcal{R}(\xi_t, L):=\left\{
(u^e(t), y^e(t)):
\begin{gathered}
\text{there exist } \alpha(t), u(t), y(t) \\
\text{ s.t. }
\eqref{cs1}-\eqref{cs4}\text{ hold}
\end{gathered}
\right\}.
\nonumber
\end{equation}
The optimal achievable stage cost $L$ steps in the future is defined as
\begin{equation}
l^o(\xi_t, L)=\min_{(u, y)\in\mathcal{R}(\xi_t, L)} l(u, y).
\nonumber
\end{equation}
\end{mydef}
%
%

Note that $l^o(\xi_t,L)$ can be computed based only on the available data.
The optimization problem \eqref{optprob1} is solved in a receding horizon manner, see Algorithm 1 .
For a given initial condition $\xi_0$ at time $t=0$, we choose the upper bound value of the terminal cost as $\bar{l}(0)\geq l^o(\xi_0, L)$.
After the initial time, the upper bound value $\bar{l}(t)$ is chosen as the stage cost of the optimal artificial equilibrium at the last time step.
\begin{algorithm}[htb]\label{algorithm1}
\caption{Data-driven EMPC scheme.}
\begin{algorithmic}[1]
\State At time $t=0$, choose a value of $\beta>0$, prediction horizon $L$, stage cost function $l(u, y)$, constraint sets $\mathbb{U}$ and $\mathbb{Y}$, and generate $\{u_k^d, y_k^d\}_{k=0}^{N-1}$.
For the initial condition $\xi_t$, choose $\bar{l}(0)\geq l^o(\xi_0, L)$.
\State Solve the problem \eqref{optprob1}.
\State Apply the input $u_t=u_0^*(t)$.
\State Set $\bar{l}(t+1)=l(u^{e*}(t), y^{e*}(t))$ and $t=t+1$. Go back to 2.
\end{algorithmic}
\end{algorithm}

\subsection{Closed-loop guarantees}\label{clp}
In this section, we first show recursive feasibility and constraint satisfaction of the proposed scheme.
Then, we prove that the asymptotic average performance of the closed-loop system can be made arbitrarily close to the performance of the optimal steady-state by changing the cost parameter.

We assume that the input $u^d$ is sufficiently rich to predict the behavior of the system via Theorem 1.
\begin{assum}
The input $u^d$ generating the data is persistently exciting of order $L+2n+1$.
\end{assum}

According to the constraint \eqref{cs3}, the prediction horizon $L$ must be long enough.
\begin{assum}\upshape\label{assum3}
The prediction horizon satisfies $L\geq n$.
\end{assum}

Finally, we have the following assumption.
\begin{assum}
The feasibility set for the data-driven EMPC problem with terminal equality constraint is non-empty, i.e.,
$\mathcal{F}^s\neq \emptyset$.
\end{assum}

The following theorem states that the proposed scheme guarantees recursive feasibility and constraint satisfaction.
\begin{mythm}\upshape
Suppose Assumptions 1--3 hold and consider an initial condition $\xi_0\in\mathcal{F}^s$.
If problem \eqref{optprob1} is feasible at initial time $t=0$, then
\begin{enumerate}[(i)]
\item it is feasible at any $t\in\mathbb{N}$;
\item the closed-loop trajectory satisfies the constraints, i.e., $(u_t, y_t)\in\mathbb{U}\times\mathbb{Y}$ for all $t\in\mathbb{N}$.
\end{enumerate}
\end{mythm}
\begin{pf}
We can prove recursive feasibility and constraint satisfaction following the usual argument to construct the input at time $t$ via the optimal input in the last time $t-1$, compare \cite{rawlings2017model} for the standard proof and \cite{berberich2020changing} for the case with artificial setpoint and data-driven prediction model.
$\hfill\qed$
\end{pf}


In general, the stage cost $l$ may not attain its minimum at an equilibrium.
Thus, the asymptotic average performance criterion is used to analyze the performance of EMPC schemes, compare \cite{diehl2011empc}.
Given an initial condition $\xi_0$ and the control law obtained by Algorithm~1, the asymptotic average performance is defined as
\begin{equation}
J_{\infty}^*(\xi_0)=\limsup_{T\rightarrow\infty}\frac{1}{T+1}\sum_{t=0}^{T} l(u_t, y_t).
\nonumber
\end{equation}

Now, we are ready to state our main result on the asymptotic average performance of the closed-loop system.
\begin{mythm}\upshape
Suppose Assumptions 1--3 hold and consider an initial condition $\xi_0\in\mathcal{F}^s$.
For any $\epsilon>0$, there exists $\underline{\beta}(\epsilon)$ such that for all $\beta\geq \underline{\beta}(\epsilon)$, the following properties hold:
\begin{enumerate}[(i)]
\item  the terminal stage cost is at most $\epsilon$-suboptimal w.r.t. the optimal cost $l(u^{oe}, y^{oe})$, i.e.,
$$l(u^{e*}(t), y^{e*}(t))\leq l(u^{oe}, y^{oe})+\epsilon, \forall t\geq 0;$$
\item the asymptotic average performance of the closed-loop system obtained by Algorithm 1 is bounded as
$$J_{\infty}^*(\xi_0)\leq l(u^{oe}, y^{oe})+\epsilon.$$
\end{enumerate}
\end{mythm}
\begin{pf}
By the Fundamental Lemma in Theorem 1, constraint \eqref{cs1} provides an exact model for the LTI system \eqref{system}.
Further, the input-output cost $l(u_k(t), y_k(t))$ in \eqref{ob1} can be rewritten as an input-state cost
$$\tilde{l}(u_k(t), x_k(t))=l(u_k(t), Cx_k(t)+Du_k(t)).$$
Thus, the closed-loop under Algorithm~1 is equivalent to the closed-loop under the model-based EMPC scheme by \cite{fagiano2013generalized} with stage cost $\tilde{l}$.
We can apply Theorem 2 by \cite{fagiano2013generalized} to prove the above statement, noting that the assumptions hold in the considered LTI setup with continuous stage cost function and compact input constraints.
$\hfill\qed$
\end{pf}

According to Theorem 2 (ii), the asymptotic average performance is no worse than the cost of the optimal equilibrium $(u^{oe}, y^{oe})$ plus a tolerance $\epsilon$, which can be made arbitrarily small by choosing a sufficiently large value of $\beta$.
We can compute $\underline{\beta}(\epsilon)$ based on two $\mathcal{K}_\infty$ functions regarding the stage cost function and dynamics of the system, compare \cite{fagiano2013generalized} for details.
Even though Theorem 3 assumes the initial condition to lie in the feasibility region of the scheme with terminal equality constraints, i.e., in a suitable region around $(u^{oe}, y^{oe})$, the implementation of Algorithm 1 does not require knowledge of $(u^{oe}, y^{oe})$.
This is an advantage compared to the existing data-driven MPC scheme by \cite{berberich2020changing}.
Finally, we note that the condition $\xi_0\in \mathcal{F}^s$ can be relaxed under suitable assumptions when using a slightly more sophisticated EMPC scheme, see~\cite[Algorithm 3]{fagiano2013generalized} for details.

\section{Data-driven EMPC with unknown cost function}\label{unknown}
There are many practical applications, in which the cost function that is to be minimized is not known exactly or unknown.
For example, in robotic trajectory planning, the agent has little prior information about the cost function of entering the obstacle areas.
It only knows the value of cost when such a situation happens.
Recent works have started to consider unknown cost function in the context of learning-based and data-driven control, e.g., \cite{nonhoff2022online, manzano2021oracle, gros2020data}.
For example, \cite{gros2020data} propose a data-driven EMPC scheme to approximate the value function and control policy in reinforcement learning by suitable adaptation of parameters including the cost function.
In this section, we propose a data-driven EMPC scheme to handle unknown cost functions using the Fundamental Lemma.

We assume that the stage cost function is in a linear form with unknown parameters.
The proposed scheme in this section can be applied to any stage cost function that can be parameterized linearly.
\begin{assum}
The stage cost function is a linear function $l(u, y)=l_u^\top u+l_y^\top y$, where $l_u\in\mathbb{R}^m$ and $l_y\in\mathbb{R}^p$ are unknown parameters.
\end{assum}

Throughout this section, we assume that an explicit expression of $l(u, y)$ is not available, but we only have a cost trajectory $\{l^d_k\}_{k=0}^{N-1}$, which satisfies $l_k^d=l(u_k^d, y_k^d)$ for the given input-output data.

We now present the data-driven EMPC problem with unknown stage cost function and generalized terminal constraint.
At time $t$, we have an input-output-cost trajectory  $\{u_k^d,
y_k^d, l_k^d\}_{k=0}^{N-1}$ of system \eqref{system}, initial conditions $\{u_k, y_k\}_{k=t-n}^{t-1}$ and a bound $\bar{l}(t)\geq l(u^{oe}, y^{oe})$.
Based on these ingredients, we consider the following optimization problem:
\begin{subequations}\label{optprob2}
\begin{align}
J_L^*(\xi_t)=
&\min_{\alpha(t), u(t), y(t)\atop l(t), u^e(t), y^e(t)}\sum_{k=0}^{L-1}l_k(t)+
\beta l_L(t)\label{ob2}\\
\text{s.t. }\quad &
\begin{bmatrix}
    u_{[-n, L]}(t)\\
    y_{[-n, L]}(t)\\
    l_{[-n, L]}(t)
\end{bmatrix}=
\begin{bmatrix}
    H_{L+n+1}(u^d)\\
    H_{L+n+1}(y^d)\\
    H_{L+n+1}(l^d)
\end{bmatrix}\alpha(t),\label{ccs1}
\\
&\label{ccs2}
\begin{bmatrix}
u_{[-n, -1]}(t)\\
y_{[-n, -1]}(t)
\end{bmatrix}=
\begin{bmatrix}
u_{[t-n, t-1]}\\
y_{[t-n, t-1]}
\end{bmatrix},
\\
&\begin{bmatrix}\label{ccs3}
    u_{[L-n, L]}(t)\\
    y_{[L-n, L]}(t)
\end{bmatrix}=
\begin{bmatrix}
    u_{n+1}^e(t)\\
    y_{n+1}^e(t)
\end{bmatrix},
\\
&(u_k(t), y_k(t))\in \mathbb{U}\times\mathbb{Y}, \forall k\in\mathbb{I}_{[0, L]},\label{ccs4}\\
&l_L(t)\leq \bar{l}(t).\label{ccs5}
\end{align}
\end{subequations}
In constraint \eqref{ccs1}, we predict an input-output trajectory along with the corresponding cost values based on one input-output-cost trajectory, which is generated by a persistently exciting input sequence.
Compared to problem \eqref{optprob1}, the main difference is that problem \eqref{optprob2} does not require the knowledge of stage cost function explicitly, but uses the Fundamental Lemma to predict it.
The optimization problem can be solved in a receding horizon manner as in Algorithm 1.

The following theorem states that the proposed scheme with unknown cost function has the same solution as the scheme in Section 3.
\begin{mythm}
If Assumptions 4 holds, then the optimal solutions of problem \eqref{optprob1} and \eqref{optprob2} coincide.
\end{mythm}
\begin{pf}
The stage cost can be regarded as another output of the system since it is a linear function of the input and output.
To be precise, we have
\begin{equation}\nonumber
\begin{aligned}
    l(u_k, y_k)
    =&l_u^\top u_k+l_y^\top y_k\\
    =&l_u^\top u_k+l_y^\top(Cx_k+Du_k)\\
    =&l_y^\top C x_k+(l_u^\top+l_u^\top D)u_k.
\end{aligned}
\end{equation}
We can denote the extended output $\bar{y}_k=\begin{bmatrix}y_k\\l(u_k,y_k)\end{bmatrix}$ including the stage cost by
\[\bar{y}_k=\bar{C}x_k+\bar{D}u_k,\]
where $\bar{C}=\begin{bmatrix}
C\\
l_y^\top C
\end{bmatrix}$ and $\bar{D}=\begin{bmatrix}
D\\
l_u^\top+l_u^\top D
\end{bmatrix}$.
Applying the Fundamental Lemma to the LTI system with matrices $(A, B, \bar{C}, \bar{D})$, and noting that the initial conditions in \eqref{ccs2} imply $$l_{[-n, -1]}(t)=\begin{bmatrix}l(u_{t-n}, y_{t-n})\\ \vdots \\ l(u_{t-1}, y_{t-1})\end{bmatrix},$$
we conclude that the cost predictions in problems (2) and (3) are equivalent. Since all other terms are identical as well, this concludes the proof.
$\hfill\qed$
\end{pf}
\begin{remark}
It follows from Theorem 4 that all theoretical properties of Algorithm 1 derived in Section 3 remain true for the EMPC scheme based on problem \ref{optprob2}.
\end{remark}

\section{Simulation}\label{simulation}
In this section, we apply the proposed data-driven EMPC scheme to the linearization of a chemical reactor considered by \cite[Chapter 3.4]{faulwasser2018enmpc}.
The nonlinear system from \cite{faulwasser2018enmpc} is linearized at the equilibrium $x^e=(0.3575, 0.0580, 0.1)^\top$ and $u^e=0.1$, and discretized with a sampling time $T_s=0.1$ sec.
The linearized dynamics of the system are given by
\begin{equation}\nonumber
\begin{aligned}
x_{k+1}&=\begin{bmatrix}
0.7438 & 0 &-3.1180 \\
0.0267 &0.9048 &0.4728\\
0 &0 &0.9048
\end{bmatrix}x_k+
\begin{bmatrix}
-0.1666\\
0.0253\\
0.0952
\end{bmatrix}u_k,\\
y_k&=\begin{bmatrix}
1 &0 &0\\
0 &1 &0
\end{bmatrix}x_k.
\end{aligned}
\end{equation}
The system matrices are unknown, but an input-output trajectory $\{u_k^d, y_k^d\}_{k=0}^N$ of length $N=100$ is available, where the input is chosen uniformly from the unit interval $[-1, 1]$.
This trajectory is used to predict future trajectories in the proposed data-driven EMPC scheme.
The stage cost function is $l(u, y)=-y_{2}$ and the control goal is to minimize the stage cost.
Further, we impose constraints on the input and output as $\mathbb{U}=[-3, 3]$ and $\mathbb{Y}=[-5, 5]^2$.

We choose $L=15$ as the prediction horizon.
In general, for a given input-output trajectory, there are infinitely many vectors $\alpha(t)$ satisfying \eqref{cs1}.
To encourage a small norm of $\alpha(t)$, the objective function includes an $\ell_1$-norm penalty in the form of $10^{-2}\cdot \|\alpha(t)\|_1$.
The utility of such a regularization was analyzed by \cite{berberich2021guarantees} for a robust data-driven MPC scheme.
While \cite{berberich2021guarantees} considered a quadratic regularization $\|\alpha(t)\|_2^2$, we choose the $\ell_1$-norm  since the stage cost in this example is linear.

\begin{figure}
\centerline{\includegraphics[scale=0.65]{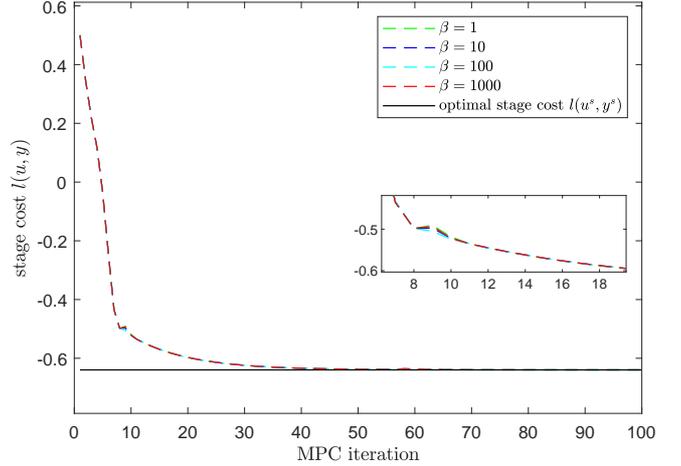}}
\caption{Stage cost of the closed-loop trajectory when using $\beta=1, 10, 100, 1000$.}\label{figure}
\end{figure}

Figure \ref{figure} illustrates the closed-loop performance, i.e., the stage cost $l(u_t, y_t)$, at each time choosing different values of $\beta=1, 10, 100, 1000$.
Based on the state-space model, we can calculate the optimal equilibrium of the system, i.e.,  $y^{oe}=(-5, 0.6396, 0.3899)$ and $u^{oe}=0.3899$, and the optimal stage cost $l(u^{oe}, y^{oe})=-0.6396$.
The closed-loop under the proposed EMPC scheme converges to the optimal stage cost.
Comparing different values $\beta$, it shows that the precise value does not have a large influence on the closed-loop performance.
The MPC scheme proposed by \cite{berberich2020changing} can also be applied for this example, but it requires a priori knowledge of the optimal equilibrium.

\section{Conclusion}\label{conclusion}
In this paper, we presented a data-driven EMPC scheme with generalized terminal constraint.
The scheme only uses input-output measurements to predict future trajectories and does not require the knowledge of an explicit model.
We proved recursive feasibility, constraint satisfaction and asymptotic average performance of the closed-loop.
Further, a novel data-driven EMPC scheme to handle unknown cost functions was proposed.
A numerical example showed the effectiveness of the proposed method.
Compared with the existing data-driven MPC scheme by \cite{berberich2020changing}, our scheme does not require the knowledge of the optimal equilibrium.
There are multiple directions for future research, which includes a theoretical and practical comparison between the proposed approach and model-based EMPC based on an identified model, see~\cite{krishnan2021direct} for open-loop results in case of a convex stage cost.
Besides, extending the results in this paper to data-driven EMPC with noisy data is another interesting future research topic, possibly using insights from model-based robust EMPC \citep{bayer2014tube}.


\bibliography{ifacconf}             








\end{document}